\documentclass{article}

\usepackage[english]{babel}

\usepackage[letterpaper,top=2cm,bottom=2cm,left=3cm,right=3cm,marginparwidth=1.75cm]{geometry}

\usepackage{amsmath}
\usepackage{graphicx}
\usepackage[colorlinks=true, allcolors=blue]{hyperref}

\title{Strategies for the Integration of quantum networks for a future quantum internet}
\author{M.I. Garc\'ia-Cid$^{1,2}$,
L. Ortiz$^{1}$, J. S\'aez$^{1}$, and V. Mart\'in$^{1}$ \\
$^{1}$Center for computational Simulation, Universidad Polit\'ectica de Madrid, Madrid, Spain \\
$^{2}$Indra Sistemas, S.A., Av. de Bruselas, 35, 28108, Madrid, Spain
}

\begin{document}
\maketitle

\begin{abstract}
The great scientific and technological advances that are being carried out in the field of quantum communications, accompanied by large investment programs such as EuroQCI, are driving the deployment of quantum network throughout the world. One of the final long-term objectives is to achieve the development of a quantum internet that provides greater security in its services and new functionalities that the current internet does not have. This article analyzes the possible integration strategies of already deployed networks or in the process of being deployed in order to reach a future global quantum network. Two strategies based on the SDN paradigm are proposed, based on a hierarchical controller scheme and on a distributed model. Each of these approaches shows pros and cons and could be applicable in different use cases. To define these strategies, the most relevant deployments of quantum communications networks carried out to date has been analyzed, as well as the different approaches for a quantum network architecture and topology, and the various proposed definitions of what quantum internet is and what are the components that would make it up in an ideal scenario. Finally, several detected opportunities and challenges regarding security and technological aspects are presented.

\end{abstract}

\section{Introduction}
\label{sec:introduction}
Quantum communications have exponentially evolved since the first testbeds that were deployed. This evolution has not only been in size but also in the underlying technologies, with new protocols and some industrialized devices. The original networks, i.e. DARPA \cite{DARPA}, UQCC \cite{UQCC}, were ad hoc dark fiber terrestrial deployments that had between $6$ and $10$ nodes and were capable of generating around $81.7~kbps$ over $45~km$.

Currently, terrestrial networks have been expanded, mainly in China, deploying more than 700 fibre-based quantum links \cite{Sat2}, with similar key rates and maintaining a dark fiber infrastructure model dedicated exclusively to quantum communications. Due to the proven feasibility of implementing these technologies covering increasingly greater distances, two major challenges have been overcome, opening the doors to real quantum communication infrastructures (QCI) and not just research-oriented deployments. The first challenge is the deployment of quantum links integrated in classical communication infrastructures in production, which represents a drastic improvement in terms of implementation costs. This approach has been implemented in \cite{MadQCI-1} using the Software Defined Network (SDN) paradigm which increases control over the network and allows a better noise management, making easier the coexistence of classical and quantum channels. The second great challenge has been the launch of Low Earth Orbit (LEO) satellites with a quantum payload \cite{Sat2} allowing two distant ground stations to be connected. This began with feasibility studies \cite{Sat0} and it has already been possible to generate key rates about 48 kbps between two cities located at a distance of 2,600 km, covering a total distance of 4,600 km \cite{Sat2}. With this achievement, together with the field tests carried out for transmitting qubits between two distant drones \cite{Drone-1}, it is possible to begin to consider theoretical architectures and feasibility studies for satellite constellations acting as key relay nodes.\\
It is clear that the technological maturity has increased much in terrestrial and spacial QCI based on fiber optics and LEO satellites, respectively. In addition, the availability of standards from standardization bodies such as ETSI, IEEE or ITU, allows a better integration into telco environments, even using devices from different providers. All this, makes possible the deployment of quantum communication networks (QCN) in metropolitan areas to guarantee the security of communications between, for example, company headquarters, bank branches, public organization or military buildings, scattered throughout the city.

In view of the advances that are taking place in the area of quantum communications and its potential, the quantum internet began to be talked about. This term began to be used in the late 1990s \cite{QI99-2} and has regained interest in the second decade of the 21st century.
Although elements that will build the future quantum internet and the possible functionalities that it can provide have been proposed and defined \cite{QI18}, there are still gaps when referring to the definition of a common architecture for integrating the current quantum networks that will support the future quantum internet. \\
That is why, in this article, we define and compare a hierarchical and a distributed architectural model for the integration of quantum communication networks. For that, an analysis of the previously mentioned quantum communication deployments has been performed, as well as of today's technological and security challenges that still exist and that must be solved. 

The article is structured as follows, Section \ref{sec:Isolated QComms} describes the components of an ideal scenario for a quantum network, different topologies for isolated quantum networks  on the Software-Defined Network (SDN) approach. Taking all the previous concepts into account, Section \ref{sec:QI perspective} reviews what is the perspective of the future quantum internet. Two architectural approaches based on hierarchical and distributed models for the integration of the different types of isolated networks are presented in Section \ref{sec:Integration model}. Finally, several detected opportunities and challenges regarding security and technological aspects are presented in Section \ref{sec:security}.

\section{Isolated Quantum network architectures}
\label{sec:Isolated QComms}
When implementing QCI, different strategies have been used, such as the deployment of ad-hoc dark fiber infrastructures for the exclusive use of quantum communications or the integration of quantum technologies such as QKD devices and fiber optics-based quantum links in current communication networks in production, sharing the infrastructure with the classical links. The first strategy optimizes the quantum channel for a better performance, but it implies higher deployment costs and a lower scalability and flexibility of the network. The second strategy is less expensive, more scalable, flexible and manageable, but it is affected by the noise of the classic communications and a network design optimized for classical communications with which it coexists, decreasing the quantum network performance. However, the later approach shows a more realistic choice when it comes to large network deployments, such as those being considered for future networks. Regardless of the infrastructure deployment strategy (ad-hoc or shared), a QCN is made up of a series of disruptive elements and technologies that increase security and functionalities compared to current communication networks. In the same way as the network components, the types of applicable QCI topologies are also independent of the deployment strategy in the sense that all the topologies discussed later in this section are feasible in both cases. However, the choice of one strategy or another has an impact on the operating parameters of the network.

\subsection{Components of a QCI}
All network management and control processes, synchronization of quantum devices and key distillation processes, among others, classic network elements such as switches, routers, communication channels, servers, etc. are needed. In addition to the elements of a classical network, in order to have an ideal QCI, a series of key features are needed: quantum channels (QC), quantum repeaters (QR), quantum memories (QM) and quantum nodes (QN), as defined below. This does not mean that it is strictly necessary to have all of these components in order to have an operative QCI, but rather that having all of them at the same time, it is possible to have all the capabilities and functionalities that are expected (today) from these networks and from the future quantum internet, as we will see in Section \ref{sec:QI perspective}.

\textbf{Quantum nodes (QN)}. Depending on the context in which the quantum nodes are being discussed, they can be understood from different approaches.

\begin{itemize}
    \item \textit{Approach 1 – Nodes as devices}. Quantum nodes as each of the qubit emitting/receiving devices. In this case, a pair of QKD devices are described as two quantum nodes connected through a quantum channel.
    \item \textit{Approach 2 – Nodes as locations}. Nodes such as those physical locations that have the capacity to emit and/or receive qubits. In this approach, a quantum node could have more than one emitter/receiver module that connects it with other. 
    \item \textit{Approach 3 – Nodes as relays}. Networks sometimes have intermediate active and/or passive elements that simply relay the key generated by QKD, switch the signal path or (de)multiplex it. 
\end{itemize}

In order to later present the architecture for the integration of different quantum networks, it is also convenient to distinguish a last type of node, which can be included within any of the three previous approaches, called border node or network access \cite{Ciurana}. A border node is that node within a network domain that is linked to the border node of a different one, thus enabling the communication between different network domains. From here on we will use approach 2 to refer to the nodes of a quantum network.

\textbf{Quantum channels (QC)}. Several transmission media for qubits acting as quantum channels are available: optical fibre, open air, space and water. Each of them has its own characteristics. When the transmission of photonic qubits is carried out through optical fiber, the signal loss increases exponentially with distance, but metropolitan distances with losses less than about $30dB$ are within reach using available technology. In this case, quantum channels could - under certain circumstances \cite{MadQCI-1} - share infrastructure with classical communications in order to minimize deployment costs and increase the availability of this technology. For its part, free space transmission in metropolitan areas suffers from turbulence and other atmospheric effects. Satellite transmission seems the most optimal option for long-range quantum communications since the qubits would suffer the major disturbance through 10 km of thickness corresponding to the lower atmosphere where the losses are higher, but when entering the space regime, these losses would drop. However, this type of channel is not easy to control. Low Earth Orbits are better for QKD performance, but the visibility period of the satellite is only a few minutes. Geostationary satellites provide 24/7 visibility, but the great distance means more dispersion and requires large telescopes for delivering a better performance. Finally, in recent years feasibility studies have been carried out to make quantum communications in an aquatic environment \cite{Water-3}.

\textbf{Quantum memories (QM)}. Normally, in QKD protocols, qubits are received and measured, storing their final classical value ($0$ or $1$). However, in order to have applications beyond the conventional QKD, it may be necessary to store the qubit in its quantum state, that is, before being measured. For that, quantum memories are necessary. QM are necessary components to store the entangled qubits in the nodes, preserving their quantum properties and to carry out operations (quantum gates) on them in order to transmit the quantum correlations. This is an essential step to support teleportation, a central application for the creation of the non-classical correlation that are the key ingredient to support the new capabilities afforded by a QCI. 

\textbf{Quantum repeaters (QR)}. Qubits cannot be copied as classical signals can, which is why it is a technological challenge to build a quantum repeater. Since today there is no QR, an approach based on trusted nodes has been adopted to reach greater distances. Trusted nodes start from the premise that intermediate nodes are trustworthy since the information is decrypted and re-encrypted in those intermediate nodes. In contrast, quantum repeaters do not require the intermediate or repeater nodes to be trusted and, not only they increase the distances in quantum communications, but also, they can establish entanglement between two distant nodes following the teleportation process.

\subsection{Network topologies}
\begin{figure}
\centering
\includegraphics[width=0.3\linewidth]{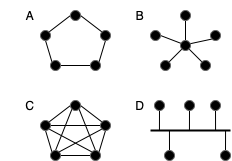}
\caption{\label{fig-topologies}Quantum network topologies corresponding to (A) ring, (B) star, (C) mesh and (D) bus.}
\end{figure}

Although an ideal QCI would be composed of all the elements described in the previous section, the reality is that with the technological maturity level of QR and QM, current networks are made up of the minimum components deployed and managed in different ways depending on the network architecture, topology chosen and the functionality expected from the network. These components are classical and quantum channels, pairs of QKD modules, optical elements such as switches, Wave-Division Multiplexers, Time-Division Multiplexers or Polarization-Division Multiplexers and common passive network components. Quantum networks can be deployed in equivalent topologies to those found in classical networks as showed in Figure \ref{fig-topologies}, being each black dot a QN (approach 2). Thus, to date QCI can be found in ring, star, mesh or bus topology or hybrid configurations of several of these topologies. Being hybrid topologies those made up of a network backbone connecting different subnetworks, each one with its own topology.

As an example, without the need for QR, a quantum network in star topology can be obtained by making use of a passive optical switch as the central element that redirect qubits to any QN applying a one-to-many approach \cite{O2M}. And the absence of QM means that the qubits must be measured upon receiving them, obtaining and storing the classical output in key management systems (KMS).

\subsection{Networks based on SDN}
\begin{figure}
\centering
\includegraphics[width=0.8\linewidth]{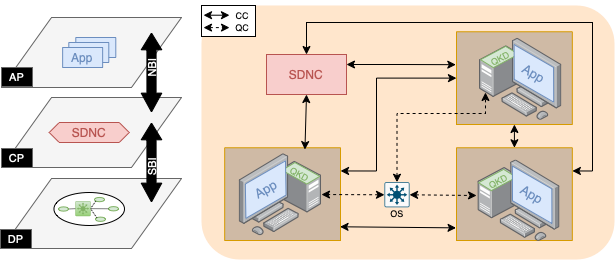}
\caption{\label{fig-SDN}SDN quantum network. (Left) Data (DP), control (CP) and application (AP) planes of the SDN model. AP-CP communicate through the North-bound interface (NBI) and CP-DP through the South-bound interface (SBI). (Right) Example of SDN quantum network with three QN in star topology and managed by the Software-defined network controller (SDNC).}
\end{figure}

QCN based on the Software-Defined Network (SDN) paradigm have been implemented and tested in the Spanish network MadQCI \cite{MadQCI-1}. In general, SDN facilitates the management of the network, both of the classical and quantum elements, and increases its flexibility and scalability, regardless of the chosen network topology. This approach allows the integration of QKD devices from different providers. In a trusted node model, each point-to-point link may use a different QKD protocol (COW, BB84, E91, GG02). The elements seen above would be located in the lower layer of the model showed in Figure \ref{fig-SDN}, together with the classical communication devices such as optical switches (OS) and satellite links to communicate different networks located at long distances. This layer corresponds to the data plane (DP) and it is where all the hardware elements and communication infrastructures are located to carry out the generation of qubits, their detection, processing and storage of the cryptographic keys in the KMS, in addition to classical traffic. It also contains the links between QN through quantum channels (QC), but a classical channel (CC) is also required for synchronization tasks and key distillation processes.

All these elements (classical and quantum) are orchestrated in the upper layer, which is an abstraction layer called the control plane (CP) where the SDN controller (SDNC) acts, in communication with the data plane through the South Bound Interface (SBI). The controller is in charge of collecting service requests from the application plane (management of notifications and requests), collecting the availability status of the intervening elements found in the DP (device and link management), elaborate the entire routing flow to be followed (network topology, routes between applications involved in the connection and security mechanisms) and orchestrate all the processes that have to be carried out to provide the requested service (status management). Although a single controller is capable of managing the entire network, to ensure the network security and resilience it is convenient to have a redundant controller. 

Finally, through the North Bound Interface (NBI), the control plane receives the service requests from the application plane (AP) and provides it with the information necessary to initiate the service between end users. The AP is also provided with cryptographic keys from the KMS whose forwarding module is allocated in the DP to encrypt the communication between end users' applications. This communication is done through the classical channels, whether they are fiber optics or wireless connections. Each quantum node is linked to the SDNC by means of an SDN Agent, which is located between the CP and the AP. Since quantum elements are not involved in the AP, the functional and non-functional requirements of this layer as well as the network policies (e.g., access policies) are almost replicable from classical network models, although depending on the QKD key generation rate of the network or its availability in the KMS. In addition to all the applications and services that currently exist in the AP, applications based in QKD-generated keys are added such as Network Function Virtualization (NFV), IPSEC, OPoT, quantum digital signatures (QDS), quantum secret sharing (QSS), quantum teleportation, byzantine agreement or quantum bit commitment, among others.
The European Telecommunications Standards Institute (ETSI) has already published a series of standards related to QCN based on SDN, where several specifications and interfaces currently in use are provided. Of particular relevance are the interfaces defined to communicate the key manager with the applications allocated in the AP defined in the ETSI GS QKD 004 and 014 standards, the interface between CP and DP defined in the ETSI GS QKD 015 standard and the interface between SDNC and a network orchestrator defined in the ETSI GS QKD 018 standard.

\section{The quantum internet perspective}
\label{sec:QI perspective}
Although QCI are currently being deployed in isolation in different countries, there are already initiatives, such as the EuroQCI program leveraged by the European Commission, which aims to deploy or expand QCI and integrate them to build the European Quantum Communication Infrastructure \cite{EuroQCI}. This type of programs promoting large quantum communication deployments will allow the future quantum internet (QI) to become a reality. This is why it is of great importance to analyse the current QCN and to propose a series of architectures that facilitate the network integration and evolution but, at the same time, allow the management of increasingly complex and large networks.
In this section, we will briefly review what QI is and what is expected from it in order to extract the main requirements needed to define the integration models of isolated quantum networks presented in Section \ref{sec:Integration model}.

The understanding of the Internet at the classical level can be formulated in two ways \cite{Definion-Internet}:

\begin{itemize}
    \item \textbf{Statement 1}: Set of necessary hardware and software components that compose it.
    \item \textbf{Statement 2}: Services of distributed applications that it provides.
\end{itemize}

Both approaches can be transferred when defining the QI. Thus, QI is considered as a set of devices with specific functionalities, which provides a series of services. 
After the commercialization of QKD devices, some definitions of QI have been given \cite{QI18}. Apart from the definitions, some basic properties that the QI has to fulfill can be summarized as follows:

\begin{itemize}
    \item Capability to communicate quantum information between network nodes.
    \item Cover long distances using quantum repeaters, satellites, etc.
    \item Be able to generate and distribute entangled pairs between network nodes.
    \item The infrastructure must host classical communications as well.
\end{itemize}

The property of entanglement plays a central role since it will allow to carry out quantum teleportation for the generation of non-classical correlations between users, which is a functionality that has no classical equivalent. To date it has been achieved the transmission through satellite links of distance between 1600-2400 km \cite{Sat2}. However, the transmission of entangled pairs at a reasonable generation rate is still a complex task, which means that it is a technology that is still in an early research phase.

In Section \ref{sec:Isolated QComms}, we specified the basic elements that build the architecture of a full quantum internet, following Statement 1. 
Following the approach from Statement 2, four stages that make up the development of the QI can be sketched modifying the approach presented in \cite{QI18}. Each of these phases adds a new functionality while increasing the difficulty of the developments. As a summary, Stage 1 are networks based on trusted nodes in which a prepare and measure QKD protocol is established between trusted contiguous nodes in order to share a symmetric key between two final end nodes. This is the state of current QCN. Stage 2 where quantum entanglement is established between nodes and qubits can be transmitted without the need of intermediate trusted nodes. This stage must also introduces Quantum Memories to build networks in which nodes have the ability to store a qubit in a location for certain time. In Stage 3 networks have a greater temporary storage capacity that favour functionalities such as distributed computing. And, finally, in Stage 4 networks will be integrated with quantum computers.

\section{Integration models of isolated quantum networks}
\label{sec:Integration model}
Based on the network components, topologies and SDN paradigm reviewed until now, we are going to define now two possible integration models - Hierarchical and Distributed - for isolated QCI. In both approaches, we will be referring to different domains, understanding a domain as a collection of quantum nodes that are controlled by a SDNC. In this case, a domain could be equivalent to a local quantum network. 
\subsection{Hierarchical model}
\begin{figure}
\centering
\includegraphics[width=0.5\linewidth]{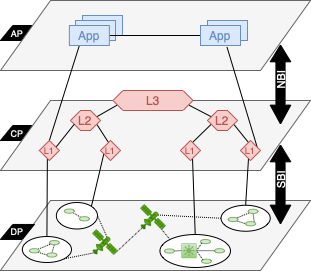}
\caption{\label{fig-hierarchical}Hierarchical model of SDN-based networks integration.}
\end{figure}

When different network domains with their own controller are integrated into a larger network, a hierarchy of controllers similar to the hierarchy of ISPs (Internet System Providers) of the current internet is established. In a specific country, the Level 1 (L1) local controller knows the status of a specific network domain and is coordinated by a higher-level regional controller (L2) which knows the status of several L1 domains. Then, the regional controllers are coordinated by a national controller (L3) which has the global network state, as showed in Figure \ref{fig-hierarchical}. More than 3 level can be set in a hierarchical model.

Establishing a hierarchy makes it possible to reduce the complexity of the operations, increase the overall network control and avoids having a single point of failure that disables the entire network by replicating the higher levels SDNC \cite{SDN-hierarchical}. On the other hand, it presents less flexibility in terms of asset reorganization.

In the diagram showed in Figure \ref{fig-hierarchical}, four isolated QCN are integrated in a hybrid topology composed of a backbone that implements a bus topology including two satellite nodes. Each subnetwork domain connected to the backbone has a different topology, specifically, two rings, a simple point-to-point link and a star topology. Each of these subnetworks has its own L1 SDNC, which are connected to their associated L2 SDNC and these to the general L3 SDNC. In this scenario, L1 SDNC do not have direct communication links between them but an interface for communicating with upper level controllers, e.g. L1 SDNC with L2 SDNC.

\begin{figure}
\centering
\includegraphics[width=\linewidth]{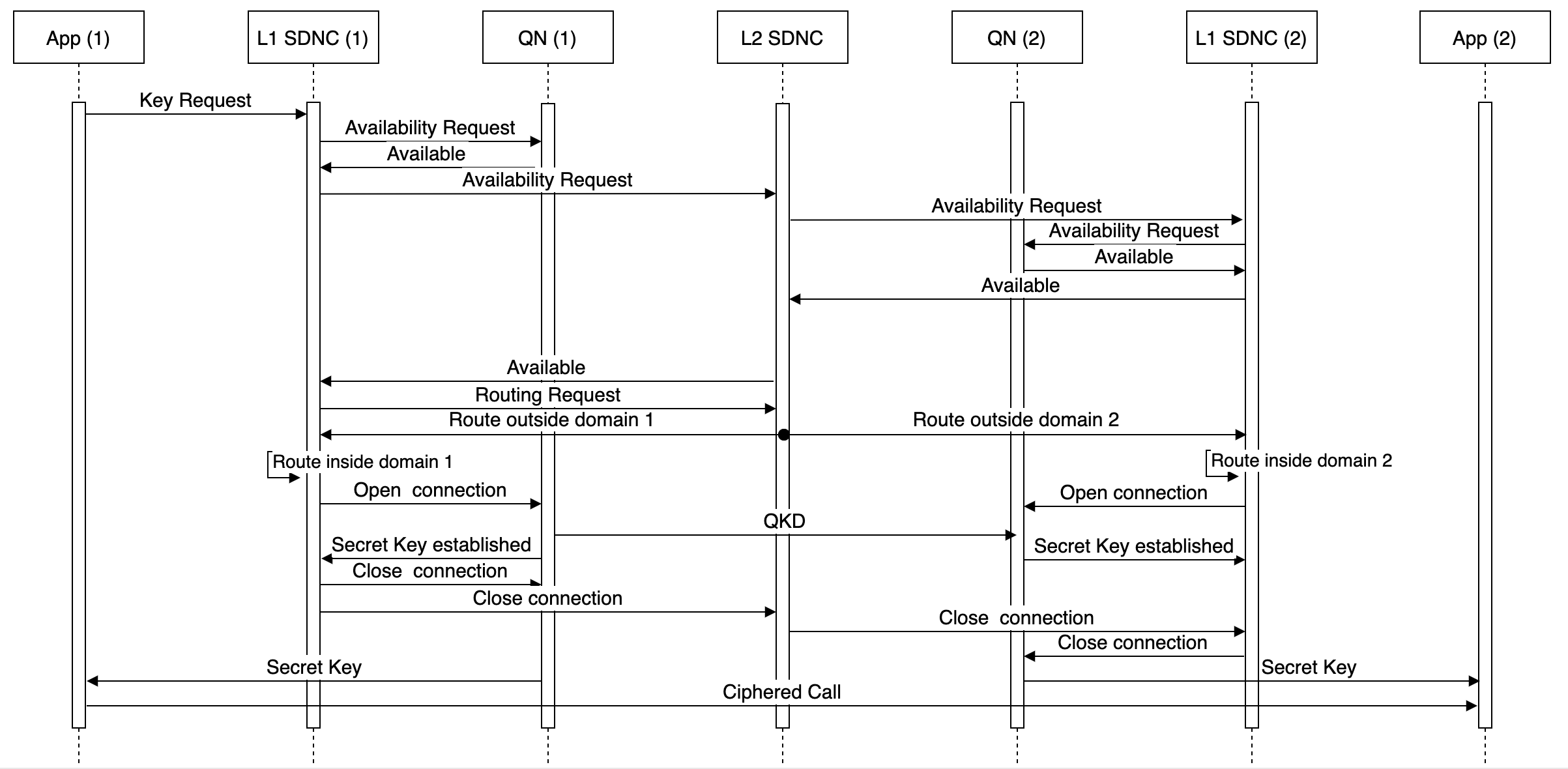}
\caption{\label{fig-videoconf-hierarchical}Sequence diagram of an encrypted videoconference stablished bewteen two network domains integrated in a hierarchical model.}
\end{figure}

As an example, Figure \ref{fig-videoconf-hierarchical} shows a sequence diagram for a use case in which two users located in different network domains and separated by a great distance want to start a videoconference application encrypted with keys generated by QKD in the context of the described hierarchical architecture.

The example involves the application (App), controller and QKD device within a quantum network on the origin (domain 1) and destination (domain 2) and a L2 SDNC. The App on domain 1 makes a request for cryptographic material to the local controller (L1 SDNC), which ask for availability to the QKD device in the QN. After confirmation, L1 SDNC detects that the destination App is out of its domain so that it communicates to L2 SDNC. L2 SDNC analyses the status of both subnetworks and communicates with L1 SDNC of domain 2. Once the availability of all the intervening elements in DP has been confirmed and the two local SDNC have been coordinated, L2 SDNC establishes the interdomain communication route between domain 1 and 2 and each L1 SDNC establishes the intradomain route inside its own domain. The connection between QN (1) and QN (2) is established and the QKD protocol is performed establishing a final secret symmetric key between both QN. The symmetric key is provided to the App at each location and, once the process is finished, L1 SDNC (1) communicates to the L2 SDNC the end of the connection and L2 SDNC to L1 SDNC (2). Finally, the end users are able to establish an encrypted videoconference.

\subsection{Distributed model}
The second SDN-based quantum communications network integration model is the distributed controller model. In this case, the management and control of the entire network is decentralized, in such a way that each domain has its own SDNC in charge of the intradomain services but at the same time each controller knows the whole network state. The different controllers communicate with each other, through an East-West Bound Interface (EWBI), to exchange network information in order to establish a path for quantum communications between two network end users belonging to different domains.
In this way, each domain can act as an isolated network or as one more segment of a large QCN.
This type of configuration favors the scalability of the network, since new domains can become part of it, as well as the robustness of the network, since it prevents single point of failures \cite{SDN-distributed}. If a network domain suffers a failure, the whole network is capable of reorganizing the routes to avoid the damaged segment. However, by not having a central controller, the amount of information that each controller has to handle increases, thus increasing the complexity of the operations. In addition, robust synchronization and prioritization mechanisms are required to avoid simultaneity in key requests that could lead to wrong key distribution.

\begin{figure}
\centering
\includegraphics[width=0.5\linewidth]{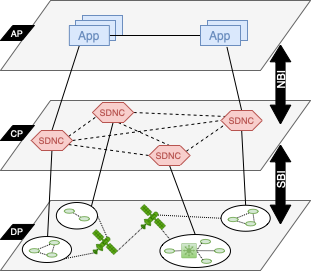}
\caption{\label{fig-distributed}Distributed model of SDN-based networks integration.}
\end{figure}

As in the previous model, four isolated QCN are integrated in the same hybrid topology composed of a backbone  bus topology, as showed in Figure \ref{fig-distributed}. Each of these subnetworks has its own SDNC, which are connected through the EWBI to SDNC from other domains, as the one defined in ETSI GS QKD 018 standard.

\begin{figure}
\centering
\includegraphics[width=\linewidth]{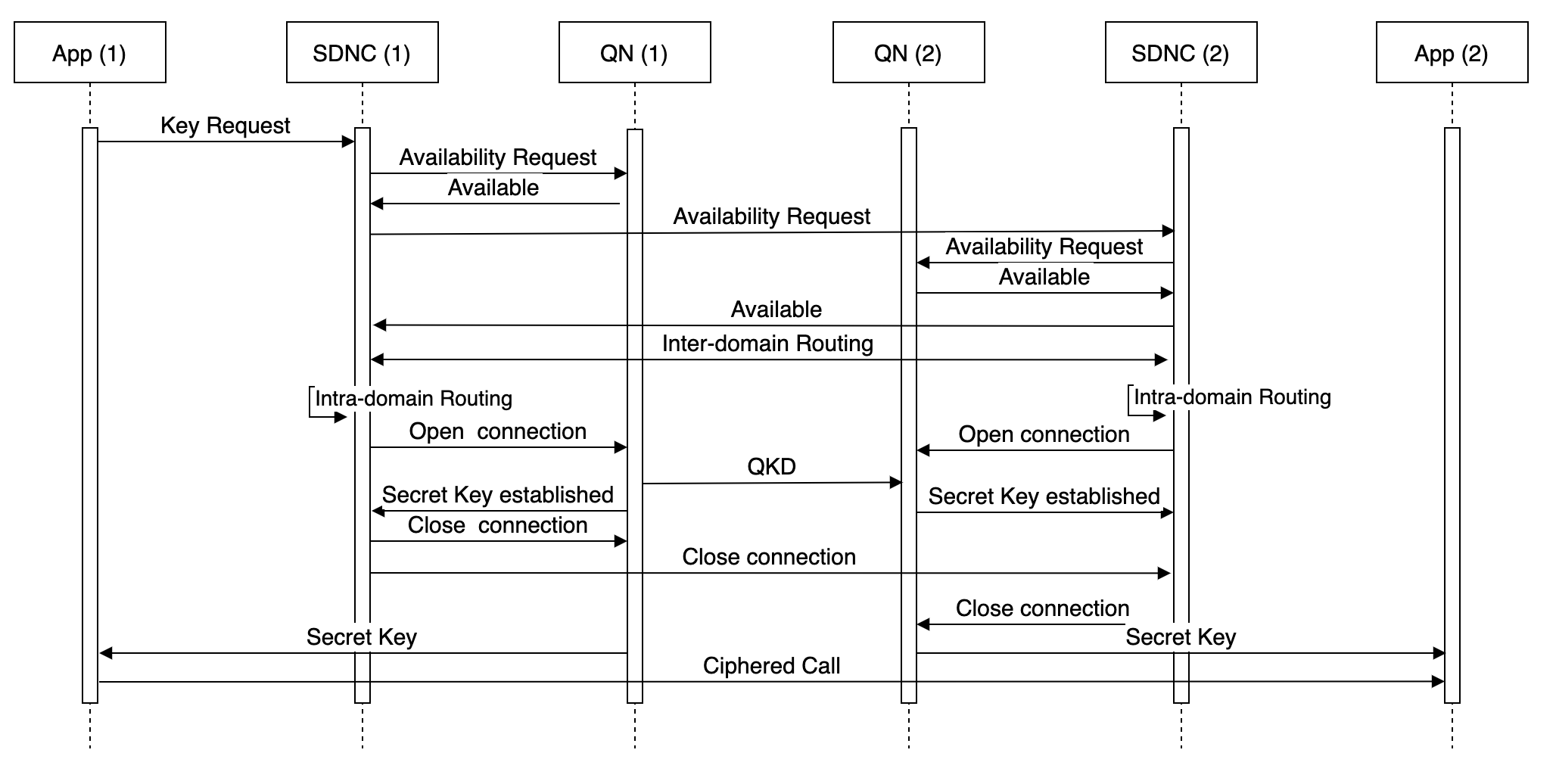}
\caption{\label{fig-videoconf-distributed}Sequence diagram of an encrypted videoconference stablished bewteen two network domains integrated in a distributed model.}
\end{figure}

For this scenario, the same use case as for the hierarchical model is depicted in Figure \ref{fig-videoconf-distributed}, but in the context of the described distributed architecture. 

The example involves the application, controller and QKD device within a quantum network on the origin (domain 1) and destination (domain 2). The App on domain 1 makes a request for cryptographic material to its associated controller, which ask for availability to the QKD device in the QN. After confirmation, SDNC from domain 1 communicates to SDNC from domain 2, which in turns performs its own availability requests with their QN. Once the availability of all the intervening elements has been confirmed, the two local SDNC coordinate and agree the communication route along the interdomain segments (outside domains 1 and 2) and each SDNC establishes the intradomain route. The connection between QN (1) and QN (2) is established and a QKD protocol is performed. Once the process is finished, the end of the connection is agreed between both SDNC. Finally, the key is provided to the App and the end users establish an encrypted videoconference.

\section{Security and technological opportunities and challenges}
\label{sec:security}
The security risk once quantum computers reach the capabilities necessary to break current public key cryptosystems such as RSA or Elliptic-curves families, could be avoided by merging classical internet infrastructures with QCI with new quantum cryptographic algorithms. This integration will also bring new functionalities that could not be achieved with classical internet in isolation. The new functionalities (those that are currently known, since others might be discovered in the future) and the advantages that the complete QI will bring over the classical internet can be summarized as follows:

\begin{itemize}
    \item High, long-term, security without computational assumptions, i.e. Information Theoretic Security (ITS). Quantum Key Distribution (QKD) makes possible the creation of shared random symmetric keys with unconditional security. 
    \item Increase in security levels by applying widely studied quantum cryptographic protocols and/or quantum assisted protocol such as quantum digital signature, oblivious transfer, quantum secret sharing, as well as hybrid approaches where classical and quantum cryptographic tools are integrated to provide cryptoagile systems.
    \item Increase in the efficiency of quantum cryptographic protocols compared to their classic counterparts, under the same operating conditions.
    \item Assisted information transmission through quantum teleportation.
\end{itemize}

These opportunities, except for teleportation capabilities, are already feasible and directly transferable into today's QCN covering metropolitan areas.
However, there are a number of elements or conditions that still require development. Throughout the previous sections we have already commented on some of the elements that still have a low technological maturity such as QR and QM, but there is also a need of implementing quantum cryptographic solution assuring the end-to-end security of the entire system. For that, the following needs to be guaranteed:

\begin{itemize}
    \item the implementation of quantum channels with measures against physical attacks or side channel attacks; 
    \item the capability to authenticate new cryptographic modules on the network for the very first time;
    \item the availability of self-sustained authentication processes during communications, for which a Wegman-Carter approach has been proposed \cite{Wegman-Carter};
    \item the capability to guarantee the identity of a user who access a service to which several users have access;
    \item the design of hybrid key management systems capable of storing and managing classical and quantum keys;
    \item the standardization and certification of quantum cryptographic modules;
    \item the standardization of hybrid cryptosystems combining classical, quantum and Post-quantum cryptographic mechanisms.
\end{itemize}

 In addition, although standards related to different SDN interfaces on quantum networks have already been published by ETSI, there is still a need of standards for a clear interface definition between controllers and between AP-CP (i.e. EWBI and NBI, respectively) in both hierarchical and distributed models but also interfaces between KMS to leverage the integration of quantum communication subnetworks to rise a full QI. All these challenges, point out the need of international programs such as EuroQCI. 

\section{Conclusions}
Terrestrial quantum communications networks have evolved enormously since their initial approaches. However, two strategies continue to prevail for the deployment of this type of networks: infrastructures based on dark fiber dedicated exclusively to quantum communications and infrastructures with the coexistence of classical and quantum communications. Regardless of the chosen strategy, networks of different types have been deployed and are capable of implementing new functionalities compared to classical networks.
Although it is true that many technological and scientific advances are still required in terms of the development of elements such as quantum repeaters or quantum memories, to have the ideal quantum internet, the reality is that it is already technologically possible to integrate quantum cryptographic protocols such as QKD to guarantee security of information in current cryptographic systems, at least in metropolitan distances.
The vision of a quantum internet based on the integration of a set of isolated communication networks leveraged by the EuroQCI program can be facilitated with an SDN perspective, thanks to the reduction in complexity in the data plane, virtualizing management and control functions of the physical elements in the CP. Based on the architecture of each network, they can be integrated through a hierarchical model of controllers that manage subnets at different levels or a distributed model where each controller is the owner of a specific domain and is capable of coordinating with neighboring domains SDNC. Each of these models has pros and cons. The hierarchical model is able to reduce the complexity of the operations and provide full network control, however, greater network control means a loss in flexibility for asset reorganization. In addition, to avoid having single points of failure, it requires SDNC redundancy at all levels. The distributed model facilitates the scalability of the network as well as its robustness, since it prevents single point of failures, but it presents a major complexity in operations, demanding robust synchronization mechanisms and prioritization policies.
In this area, a series of opportunities have been detected, such as the need to guarantee end-to-end security of the entire crypto-system, to develop hybrid key storage systems or the increase in the level of technological maturity of the elements required for a quantum internet with full capabilities. In addition, in order to carry out the integration of isolated QCN in a hierarchical or distributed model, an NBI is required between the control plane and the application plane, as well as an EWBI that details the interface between subnetwork controllers.

\section*{Acknowledgments}
This project has received funding from Indra Sistemas S.A; EIT Digital co-funded by European Institute of Innovation and Technology (EIT), a body of the European Union; and MCIN with funding from European Union NextGenerationEU (PRTR-C17.I1); funding from the Comunidad de Madrid. Programa de Acciones Complementarias, Madrid Quantum; from the European Union’s Horizon Europe research and innovation programme under the project ”Quantum Security Networks Partnership” (QSNP, grant agreement No 101114043); and the project “EuroQCI deployment in Spain” (EuroQCI, grant agreement No 101091638).

%
%
%
%

\end{document}